\documentclass[a4paper,10pt,twoside]{cpc-hepnp}

\usepackage{multicol}
\usepackage{graphicx}
\usepackage{booktabs}
\usepackage{amssymb,bm,mathrsfs,bbm,amscd}
\usepackage[tbtags]{amsmath}
\usepackage{lastpage}
\usepackage[dvipdfm]{hyperref}
\usepackage{epstopdf}

\begin{document}

\fancyhead[c]{\small Submitted to  Chinese Physics C}



\title{The performance of a double sided silicon strip detector as a transmission detector for heavy ions\thanks{Supported by National Natural Science Foundation of China (10905076, 11005127, 11075190, 11205209 and 11205221)}}

\author{%
 J L Han$^{1;1)}$\email{hanjl@impcas.ac.cn}
\quad J B Ma$^1$
\quad X G Cao$^2$
\quad Q Wang$^1$
\quad J S Wang$^1$\\
\quad Y Y Yang$^{1;3}$
\quad P Ma$^1$
\quad M R Huang$^1$S
\quad S L Jin$^{1;3}$
\quad X J Rong$^1$\\
\quad Z Bai$^{1;3}$
\quad F Fu$^1$
\quad Q Hu$^{1;3}$
\quad R F Chen$^1$
\quad S W Xu$^1$\\
\quad J B Chen$^{1;3}$
\quad L Jin$^{1;3}$
\quad Y Li$^{1;3}$
\quad M H Zhao$^{1;3}$
\quad H S Xu$^1$
}
\maketitle

\address{%
$^1$Institute of Modern Physics(IMP), Chinese Academy of Sciences(CAS), Lanzhou 730000, People's Republic of China\\
$^2$Shanghai Institute of Applied Physics(SINAP), Chinese Academy of Sciences, Shanghai 201800, People's Republic of China\\
$^{1,3}$University of Chinese Academy of Sciences, Beijing 100049, People's Republic of China\\
}

\begin{abstract}
The performance of a double sided silicon strip detector (DSSSD), used for position and energy detection of heavy ions, is reported. The analysis shows that the incomplete charge collection (ICC) and charge sharing (CS) effects of the DSSSD give rise to a loss of energy resolution, however the position information is recorded without ambiguity. Representations of ICC/CS events in the energy spectra are shown and their origins are confirmed by correlation analysis of the spectra from both junction side and ohmic side of the DSSSD.
\end{abstract}



\footnotetext[0]{\hspace*{-3mm}\raisebox{0.3ex}{$\scriptstyle\copyright$}2013
Chinese Physical Society and the Institute of High Energy Physics
of the Chinese Academy of Sciences and the Institute
of Modern Physics of the Chinese Academy of Sciences and IOP Publishing Ltd}%

\begin{multicols}{2}

\section{Introduction}
\label{intro}
The use of double sided silicon strip detectors (DSSSD) has become increasingly widespread over the past decades as manufacturing techniques and reliability have been improved\cite{hall1}. The DSSSDs with numbers of strips have been widely used in the detection of radioactive heavy ions profiting from their high performance in energy and spacial resolution, and the covering of a large solid angle at the same time\cite{lampcd,leda}. Several detector systems with DSSSD arrays have been build for the experimental researches in the fields of radioactive beam physics and nuclear astrophysics, such as CD\cite{lampcd}, LEDA\cite{leda}, MUST\cite{must} and DRAGON\cite{dragon}.

The silicon strip detectors rely on a high quality SiO$_{2}$ layer grown on the silicon\cite{kem80,kem84} to separate the finely segmented electrodes fabricated by doping the silicon. The SiO$_{2}$ separation area, i.e. the interstrip region of the detector, will induce incomplete charge collection (ICC) and charge sharing (CS) effects. These effects in large area single sided silicon strip detectors\cite{int87,dsdi} and segmented silicon sensors\cite{poe13} have been studied. This paper reports some characteristics of ICC and CS effects of a DSSSD working in transmission mode for heavy ion beams.

In this investigation, the DSSSD is used as transmission detector to detect the position and energy loss of heavy ions from an elastic scattering experiment\cite{yang13,yang132}. The origination of the ICC and CS evnets are confirmed, and the energy and position determination of these events are evaluated. This investigation is helpful for the data analysis of extracting the elastic scattering events and their position information\cite{yang13}.

Section 2 describes the DSSSD and the associated electronics briefly. Section 3 is devoted to a description and analysis of the test result with $^{241}$Am $\alpha$ source. The response of the DSSSD for highly ionising transmitting particles is discussed in Section 4. Conclusions are drawn in Section 5.

\section{Experimental procedure}
\label{sec2}
The DSSSD used for this investigation is an oxide passivated, ion-implanted, n-type silicon strip detector manufactured by EURISYS MESURES (type IPS48x48-150NxNy48). It consists of 48 aluminium strips of 0.9 mm width and 0.1 mm SiO$_{2}$ separations on junction (p$^{+}$) side, and 0.8 mm width strips with 0.2 mm separations on ohmic (n$^{+}$) side. Junction and ohmic side strips are orthogonal providing effective pixel areas of $\sim$0.72 mm$^2$. The thickness of the detector is 150 $\mu$m. A perspective drawing of the DSSSD is shown in Fig.\ref{fig_dsssd}.

\begin{center}
  \includegraphics[width=8.5cm]{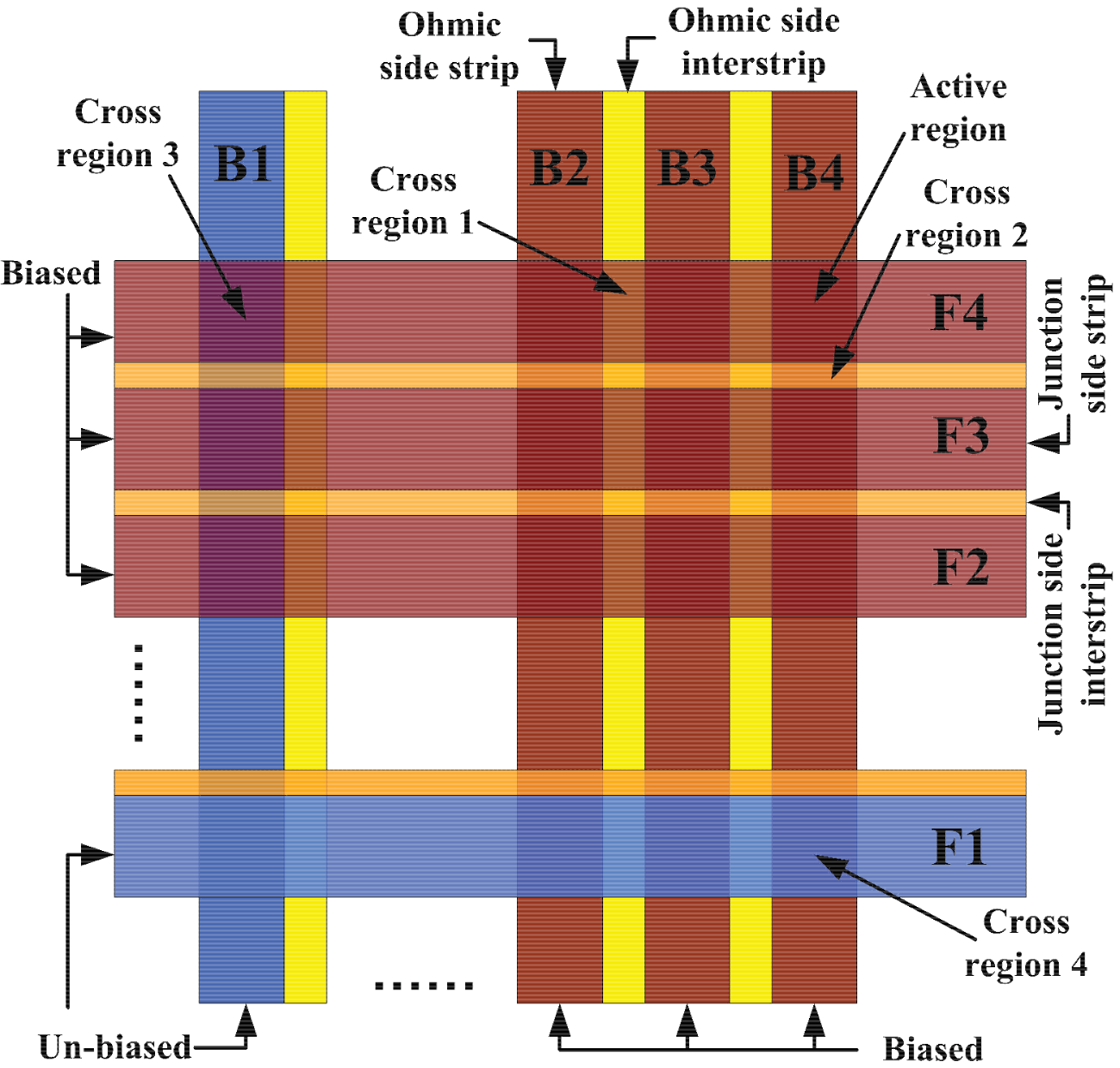}\\
  \figcaption{\label{fig_dsssd} Perspective drawing of the DSSSD, not in scale. The junction side faces to the reader. The width of junction and ohmic side strips is 0.9 mm and 0.8 mm respectively, the width of interstrip gaps is 0.1 mm and 0.2 mm respectively. The biased and un-biased strips are corresponding to the situation of test with $^{241}$Am $\alpha$ source, and all of the junction and ohmic side strips are biased for the test with heavy ion beams. Active region represents the normal working pixels of biased junction side strips cover biased ohmic side strips. Cross region 1 represents the areas of junction side strips cover ohmic side interstrips. Cross region 2 represents the areas of junction side interstrips cover ohmic side strips. Cross region 3 represents the areas of biased junction side strips cover un-biased ohmic side strips. Cross region 4 represents the areas of un-biased junction side strips cover biased ohmic side strips. See context for details.}
\end{center}

In order to study the response of the DSSSD, the entire area of junction side was illuminated by using an $^{241}$Am $\alpha$ source firstly, and then with heavy ion beams produced by the Radioactive Ion Beam Line in Lanzhou (RIBLL)\cite{sun03} with energies which can punch through the DSSSD.

Fig.\ref{fig_elec} displays the block diagram of the front-end electronics. For the test with $\alpha$ source (\textbf{offline test}) 16 adjacent strips on both sides of the DSSSD are connected to the front-end electronics. The 16 junction side strips are biased to the advised operation voltage of -20 V referring to the user manual, and the ohmic side strips are earthed. The logical OR signal of the 16 junction side strips' signals is used as the gate/trigger signal of the standard CAMAC acquisition system (ACQ). For the investigation with heavy ion beams (\textbf{online test}) all strips of the DSSSD are biased to their full depletion voltage. A 1500 $\mu$m-thick single silicon detector (SSD) with effective area of 50$\times$50 mm$^{2}$ is used to stop the penetrating particles from the DSSSD. The DSSSD and the SSD together work as a $\Delta$E-E telescope. The gate/trigger signal for the ACQ comes from the SSD. Energy spectra of the single events from each strip are recorded, along with the spectra of any pulses which occurred in coincidence in both adjacent strips.

\begin{center}
  \includegraphics[width=8.5cm]{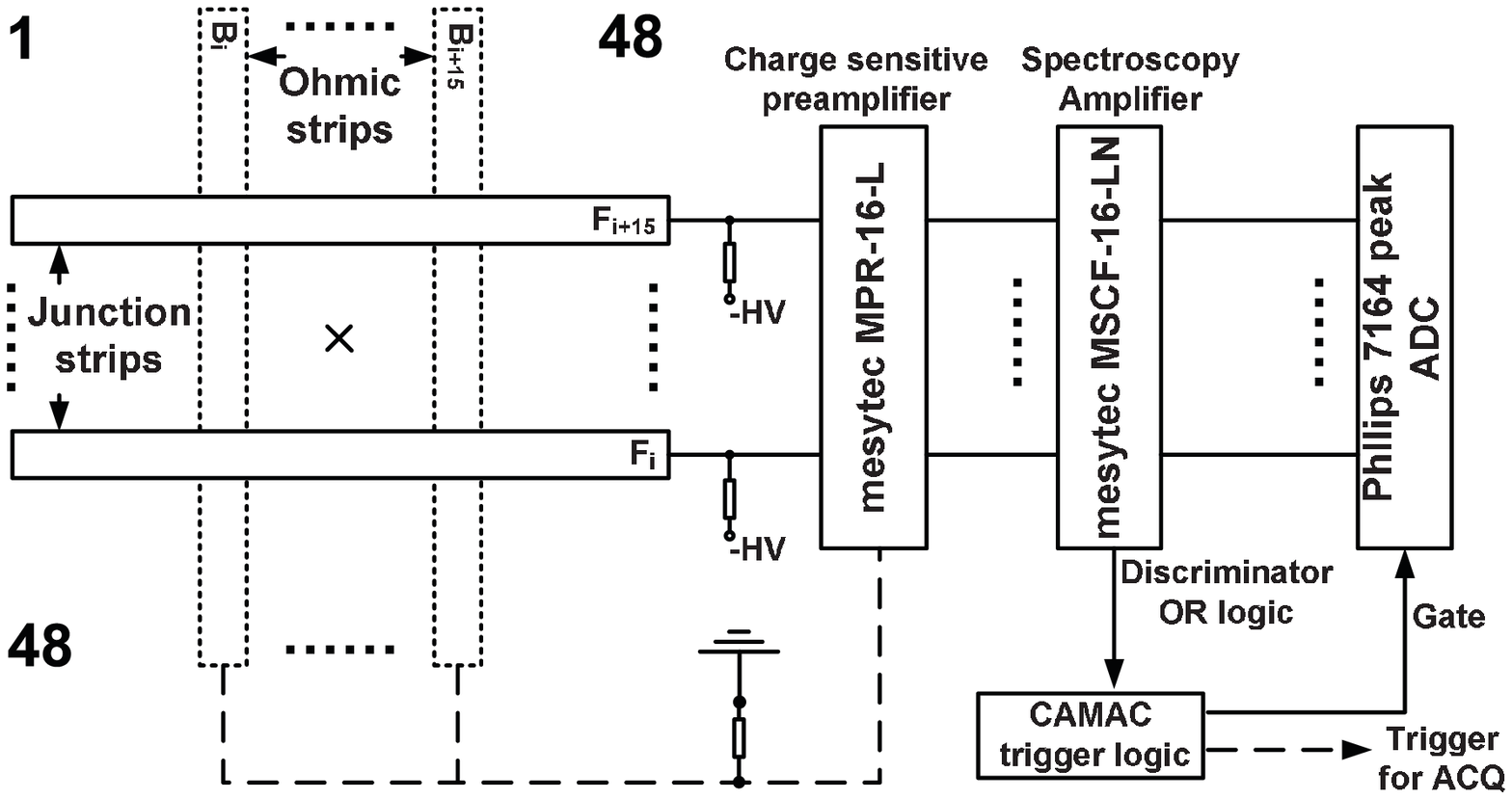}\\
  \figcaption{\label{fig_elec} Block diagram of the front-end electronics. $\times$ means the particles illuminate the DSSSD from the junction side. For the test with $^{241}$Am $\alpha$ source only 16 adjacent strips on both junction and ohmic sides are connected to the front-end electronics, for the test with heavy ion beams all strips on both sides are connected to the front-end electronics.}
\end{center}

\section{Results of offline test and analysis}

The junction side of the DSSSD is illuminated by using an $^{241}$Am radioactive source emitting 5.48 MeV $\alpha$-particles. The responses of the junction and ohmic side strips are reported in this section. ICC and CS effects, which originates from the interstrip region of the DSSSD on both junction and ohmic sides, are analyzed.

\subsection{$\alpha$-particles' energy spectrum and the CS effect of the junction side strips}
Fig.\ref{fig_falpha}(a) shows a spectrum collected from one junction side strip (for example the strip F4 in Fig.\ref{fig_dsssd}). Three sharp peaks are visible, in addition to a continuous background. The total energy peak (peak1 in Fig.\ref{fig_falpha}(a)) corresponds to the 5.48 MeV $\alpha$-particles stopped in the active region of the strip. For the active region, the corresponding junction and ohmic side strips are all connected to front-end electronics, and hence normal circuit and electric field is build in the pixel region. The full width at half maximum (FWHM) of the total energy peak, i.e. the resolution of the energy measurement taken from the junction side strip of the DSSSD, is $\sim$45 keV. The peak3 and continuous background are induced by the electron trapping effect\cite{int87,must,poe13,torr13} which results in pulse height deficits of the $\alpha$-particles stopped in the junction side interstrip regions. The ratio of the number of the counts in the peak3 and the continuous background to the counts in the whole spectrum is about 10\%. This ratio is consistent with the geometrical ratio between the area of the interstrip and strip regions on junction side.


The surprising thing is the peak2 (in Fig.\ref{fig_falpha}(a)) located near to the total energy peak. Correlation analysis of the peak2 with the ohmic side spectrum is performed. Events in the peak2 are used as restrictive cut (indicated by two up on end arrows in Fig.\ref{fig_falpha}(a)) to inspect the corresponding spectrum of ohmic side strips. The result is shown in Fig.\ref{fig_falpha}(b), it is clear that almost all of the events in the peak2 are corresponding to the low channel noise signals of ohmic side strip. The ratio of the number of the counts in peak2 to the counts of peak2 add peak1 is about 70\%. The ratio and the result of the correlation analysis lead to a conclusion that the peak2 is due to $\alpha$-particles penetrating the DSSSD from cross region 3 (see Fig.\ref{fig_dsssd}), for which only the junction strips are connected to the front-end electronics during the offline test, while the area of cross region 3 occupies 2/3$\thickapprox$70\% of the 16 junction side strips. The peak2 will disappear if all of the junction and ohmic side strips are biased.

\begin{center}
  \includegraphics[width=8.5cm]{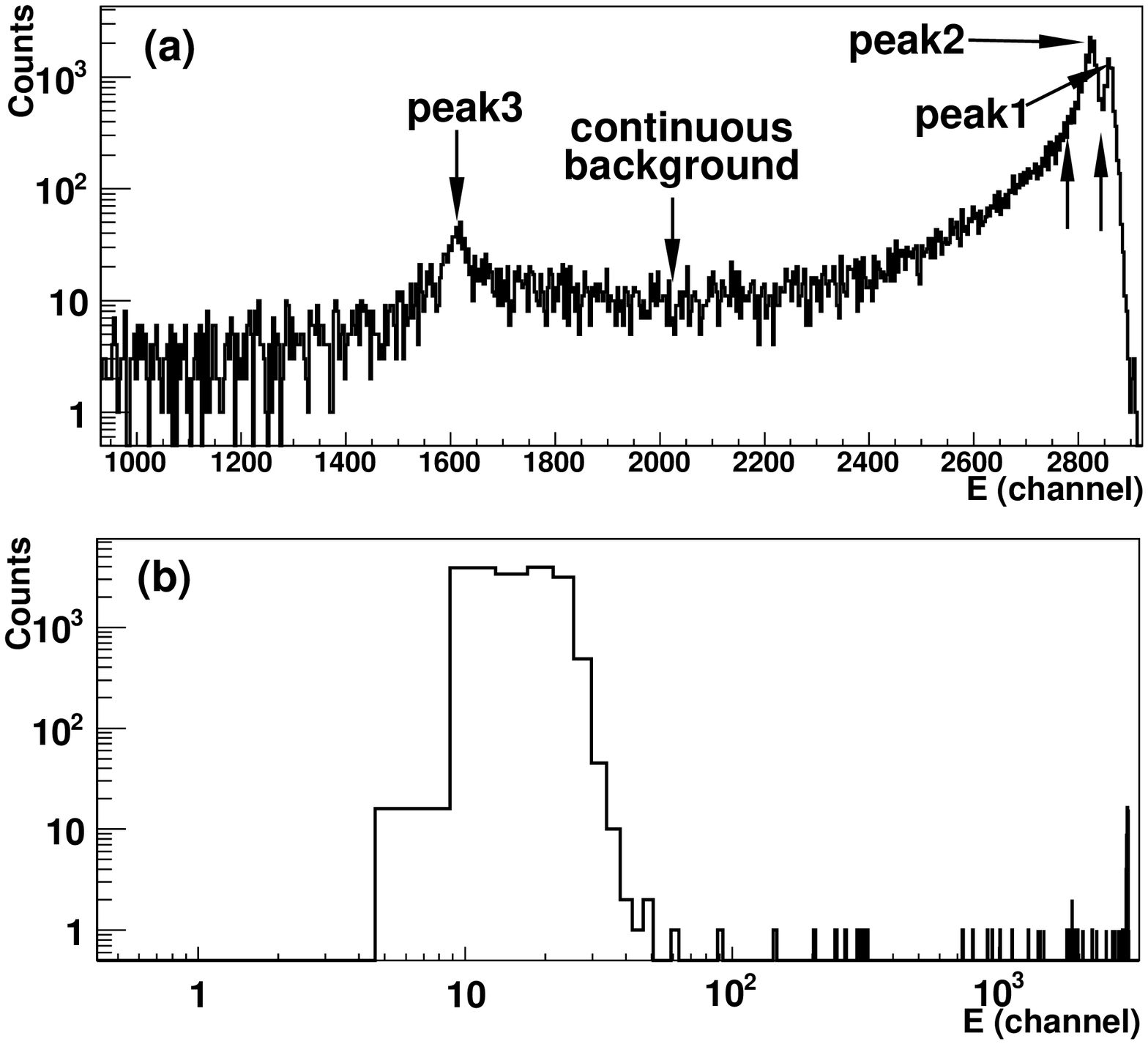}\\
  \figcaption{\label{fig_falpha} (a) $^{241}$Am $\alpha$-particles' energy spectrum from one junction side strip. (b) The energy spectrum of one ohmic side strip corresponding to the selected peak2 as indicated by the two up on end arrows in (a). See context for details.}
\end{center}

The charge correlation of two adjacent junction side strips is shown in Fig.\ref{fig_csalpha}. The negative channel values means that the output signals of those events are negative. The similar result was already observed and expatiated in Ref.\cite{int87, must} and \cite{poe13}.

\begin{center}
  \includegraphics[width=8.5cm]{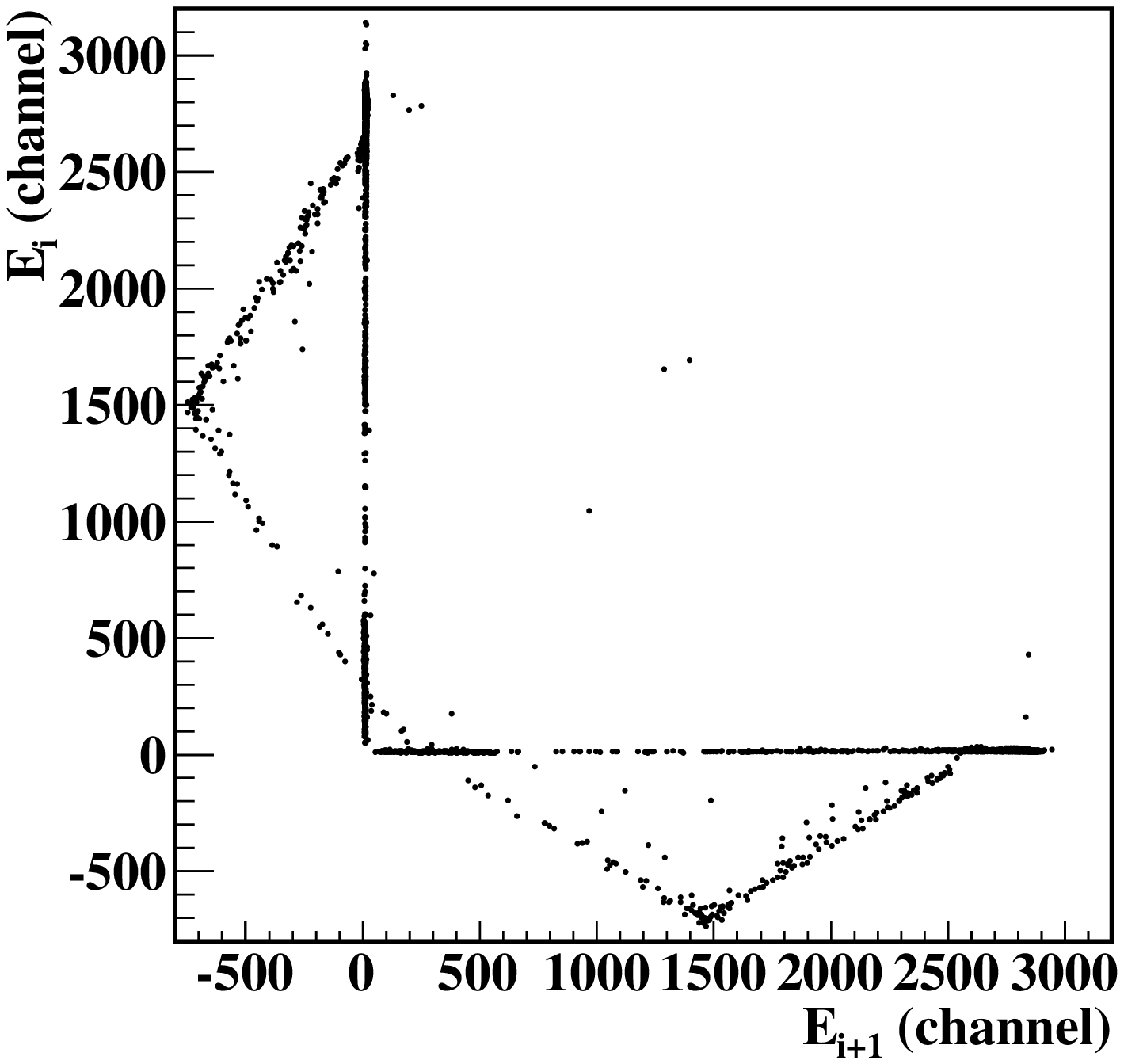}\\
  \figcaption{\label{fig_csalpha} Charge correlation between two adjacent junction side strips (strip i and i+1) while using $^{241}$Am $\alpha$ source illuminate the junction side of the DSSSD. See context for details.}
\end{center}

\subsection{$\alpha$-particles' energy spectrum and the CS effect of the ohmic side strips}
The $\alpha$-particles' energy spectrum from one ohmic side strip (for example the strip B4 in Fig.\ref{fig_dsssd}) is demonstrated in Fig.\ref{fig_bsp}(a). One sharp peak (peak1) and one minor peak (peak2) are visible, in addition to a continuous background. The peak1 with energy resolution (FWHM) of $\sim$42 keV is the total energy peak induced by the 5.48 MeV $\alpha$-particles stopping in the active region (see Fig.\ref{fig_dsssd}) of the DSSSD. The correlation analysis of peak2 with junction side spectra shows that the most of the selected events in peak2 (indicated by two up on end arrows in Fig.\ref{fig_bsp}(a)) are corresponding to the low channel noise signals of the junction side strips. The illustration of this correlation analysis is similar to Fig.\ref{fig_falpha}. This indicates that the peak2 is induced by some of those $\alpha$-particles stopped in the cross region 4 (see Fig.\ref{fig_dsssd}) of the DSSSD. The peak2 in Fig.\ref{fig_bsp}(a) seems much more minor than that in Fig.\ref{fig_falpha}. This is because that the junction side strips corresponding to the cross region 4 (see Fig.\ref{fig_dsssd}) are not biased and only those events which caused accidental coincident of the ohmic side signals with the noise signals from the junction side are recorded. The peak2 will disappear if all of the junction and ohmic side strips are biased.


The counts in the continuous background accounts for $\sim$30\% of the counts of the whole spectrum. There are two sources for the continuous background events. One source is the CS events as shown in Fig.\ref{fig_bsp}(b), the other is the particles stopping in the junction side interstrip region, which induce the electron trapping effect and hence result in incomplete electron collection of the ohmic side strips\cite{int87}. This indicates that the cause of the continuous background is attributed to the particles stopping the interstrip area of both junction and ohmic sides. The summation of the area of the interstrip region on both junction and ohmic sides occupies $\sim$30\% of the area of the DSSSD, and this percentage is consistent with the ratio of the number of counts in the continuous background to the counts in the whole spectrum.

The scatter plot of the charge correlation between two adjacent ohmic side strips is shown in Fig.\ref{fig_bsp}(b). A typical charge sharing line at the constant E$_{i}$+E$_{i+1}$ = E$_{tot}$ is observed, which corresponds to the $\alpha$-particles stopping in the cross region 1 (see Fig.\ref{fig_dsssd}) of the DSSSD. The number of the counts of the CS events occupies $\sim$20\% of the counts of the whole spectrum. This is consistent with the geometrical ratio between the area of the ohmic side interstrip and strip regions. These CS events contribute 2/3 of the counts of the continuous background as shown in Fig.\ref{fig_bsp}(a), and the other 1/3 is the contribution of these particles stopping in the junction side interstrip region.

\begin{center}
  \includegraphics[width=8.5cm]{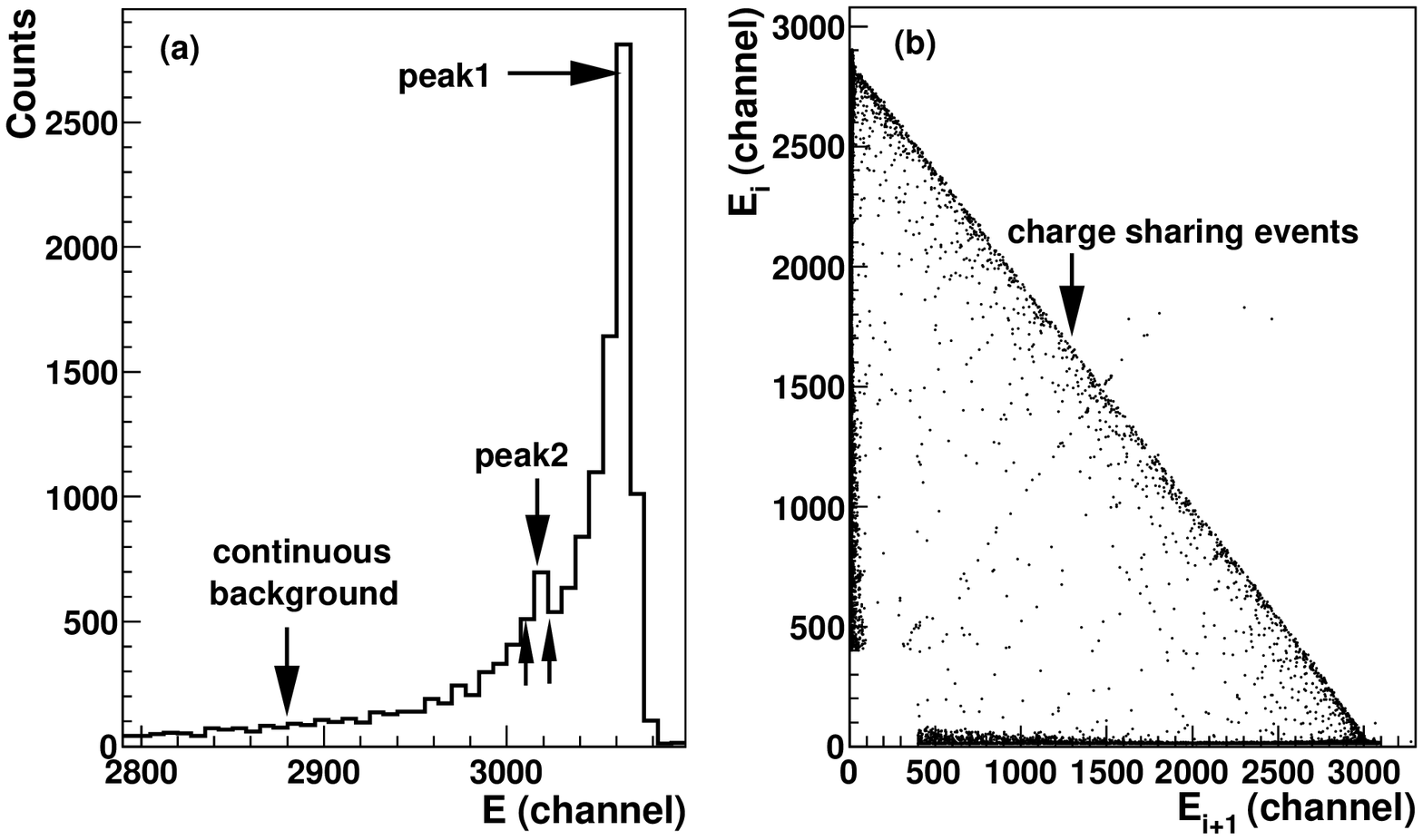}\\
  \figcaption{\label{fig_bsp} (a) $^{241}$Am $\alpha$-particles' energy spectrum from one ohmic side strip, the two up on end arrows indicates the selected range of peak2 that used in the correlation analysis of the corresponding ohmic and junction side spectra. (b) Charge correlation between two adjacent ohmic side strips (strip i and i+1), the charge sharing (CS) events are presented as a line at constant E$_{i}$+E$_{i+1}$ = E$_{tot}$. See context for details.}
\end{center}

\section{Results of online test and analysis}

Heavy ion beams with energies which can punch through the DSSSD are used to investigate the response of the detector working in transmission mode. The responses of the entrance face (junction side) and exit face (ohmic side) strips are reported and analyzed in this section.

In this test, the DSSSD and the SSD compose a $\Delta$E-E telescope used for particle identification. The experimental setup and the production of the heavy ion beams were reported in details in Ref.\cite{yang13}.

\subsection{Results of online test}
Fig.\ref{fig_chsh} shows a scatter plot of the charge correlation between two adjacent strips of the entrance face (a) and exit face (b) respectively. The CS events of the ohmic side strips show the same pattern as the result of the test with $\alpha$ source, i.e. a line at constant E$_{i}$+E$_{i+1}$ = E$_{tot}$ is obtained. The surprising thing is the entrance face strips' CS events which gives some raised events (see Fig.\ref{fig_chsh}(a)) instead of a beeline. And this will be interpreted in the next subsection.

\begin{center}
  \includegraphics[width=8.5cm]{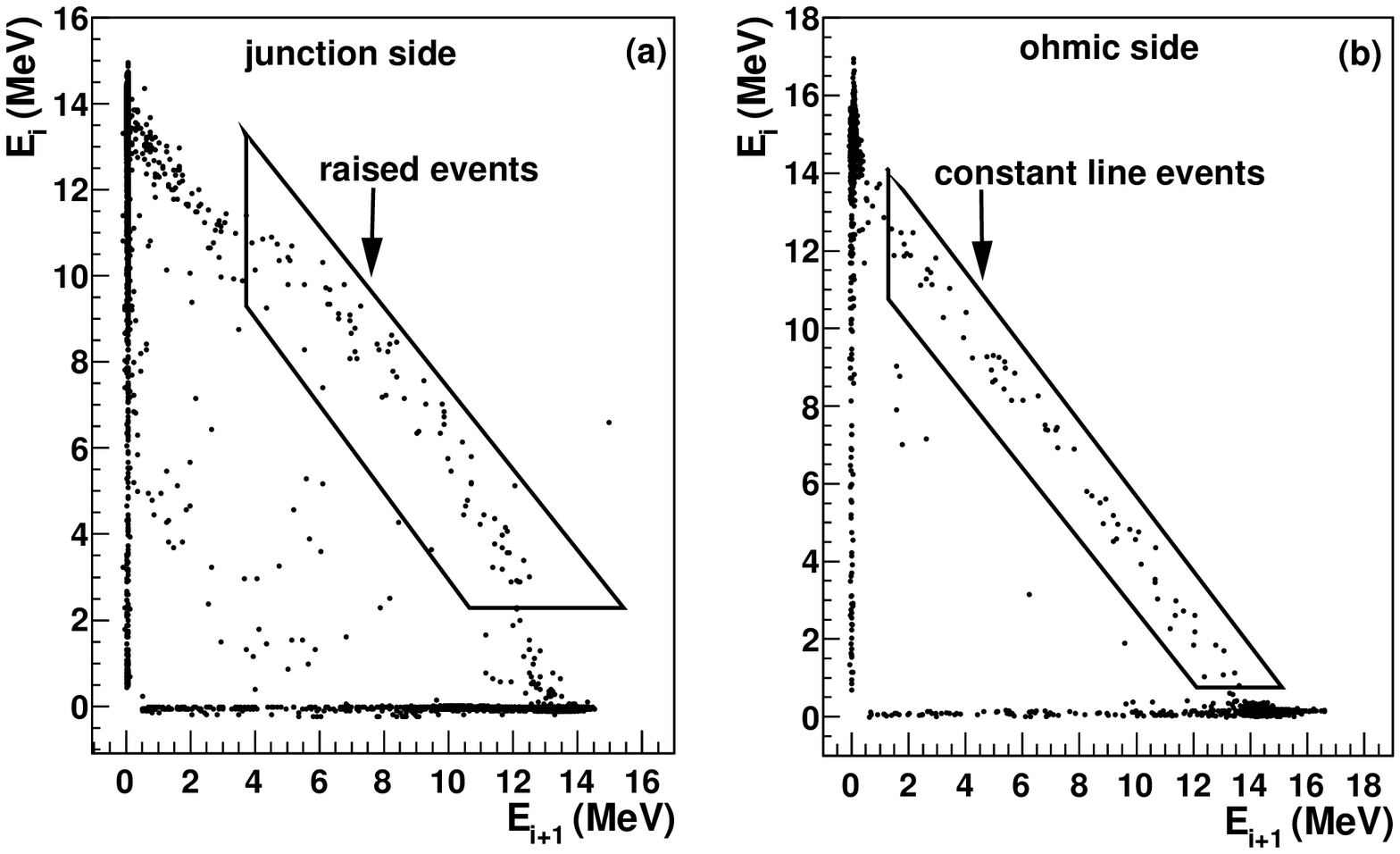}\\
  \figcaption{\label{fig_chsh} Charge correlation between two adjacent strips (strip i and i+1) of entrance face (a) and exit face (b) in the test with heavy ion beams. Some of the charge sharing events are encircled by graphical cuts, and marked as raised events in (a) and constant line events in (b). These 2 graphical cuts are examples of restrictive cuts used for correlation analysis of the $\Delta$E spectra from entrance and exit face strips. There are 47 of this kind graphical cuts for the 48 entrance and exit face strips respectively. See context for details.}
\end{center}

Fig.\ref{fig_fdee}(a) shows a $\Delta$E-E scatter plot which uses one junction side strip signal as the $\Delta$E. Several absolutely separated belts accompanied with some contaminant are visible. Fig.\ref{fig_fdee}(b) shows a $\Delta$E spectrum of the selected events in the graphical cut cut\_ELf in Fig.\ref{fig_fdee}(a). The origination of the continuous background and the peak2, i.e. the contaminant of the $^{7}$Be belt, will be expatiated infra.

\begin{center}
  \includegraphics[width=8.5cm]{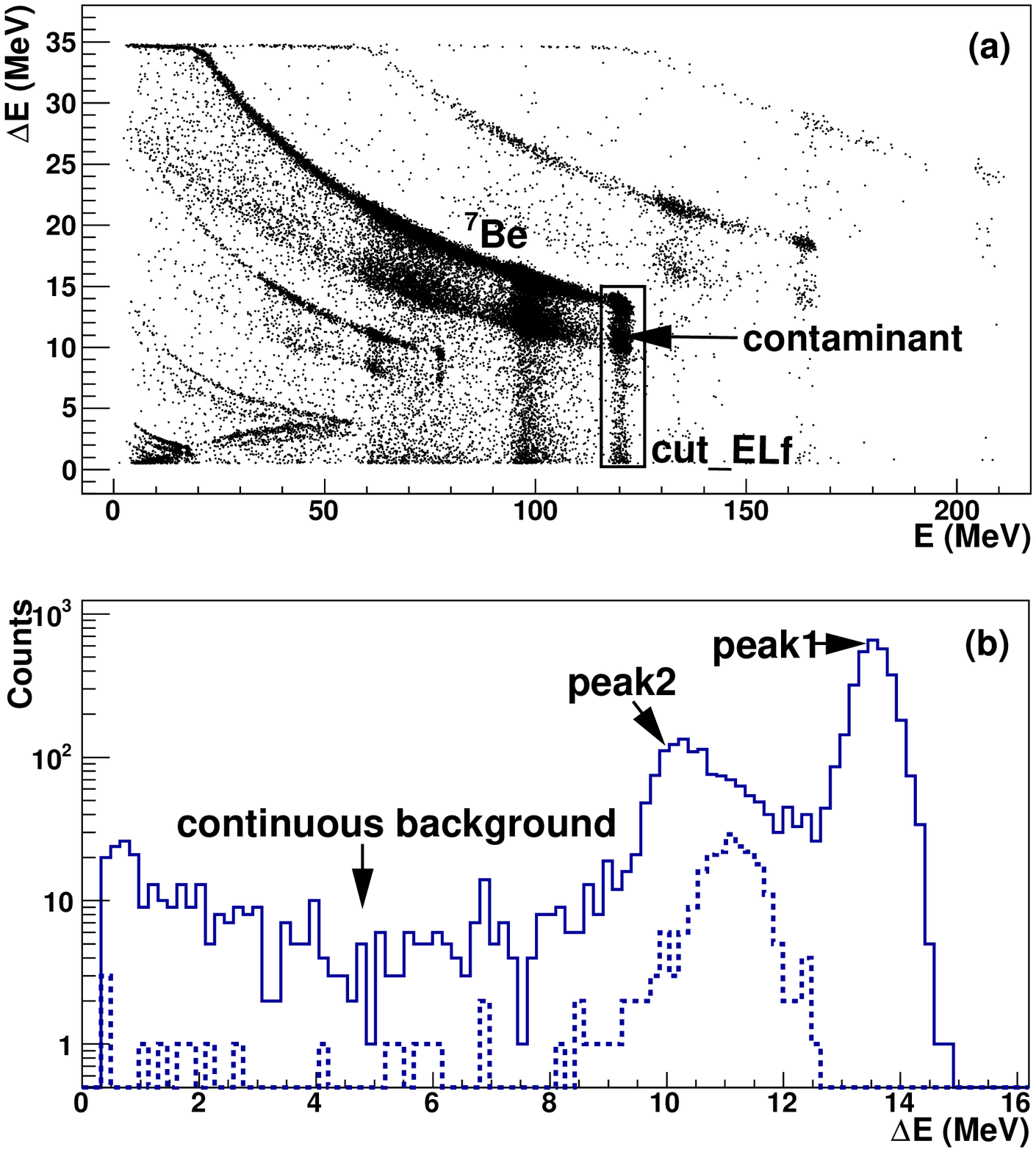}\\
  \figcaption{\label{fig_fdee} (a) $\Delta$E-E plot of the telescope. $\Delta$E is from one entrance face strip of the DSSSD, E is from the SSD. The graphical cut named cut\_ELf encircles some events used for the analysis. Some events with lower $\Delta$E energies in the graphical cut cut\_ELf are marked as contaminamt, which represents these events under each of the element belts. (b) Solid line is the $\Delta$E spectrum obtained by projecting the selected $^{7}$Be events in the graphical cut cut\_ELf to the Y-axis, and dashed line is the same $\Delta$E spectrum but some restrictive graphical cuts from exit face strips as shown in Fig.\ref{fig_chsh}(b) are imposed. See context for details.}
\end{center}

Fig.\ref{fig_bdee} is similar to Fig.\ref{fig_fdee}, but the $\Delta$E is obtained from one ohmic side strip.
Compared with Fig.\ref{fig_fdee}(b), there is only one sharp peak in Fig.\ref{fig_bdee}(b) which shows the exit face $\Delta$E spectrum, but some upper events appear. The cause of these upper events will be explained infra.

\begin{center}
  \includegraphics[width=8.5cm]{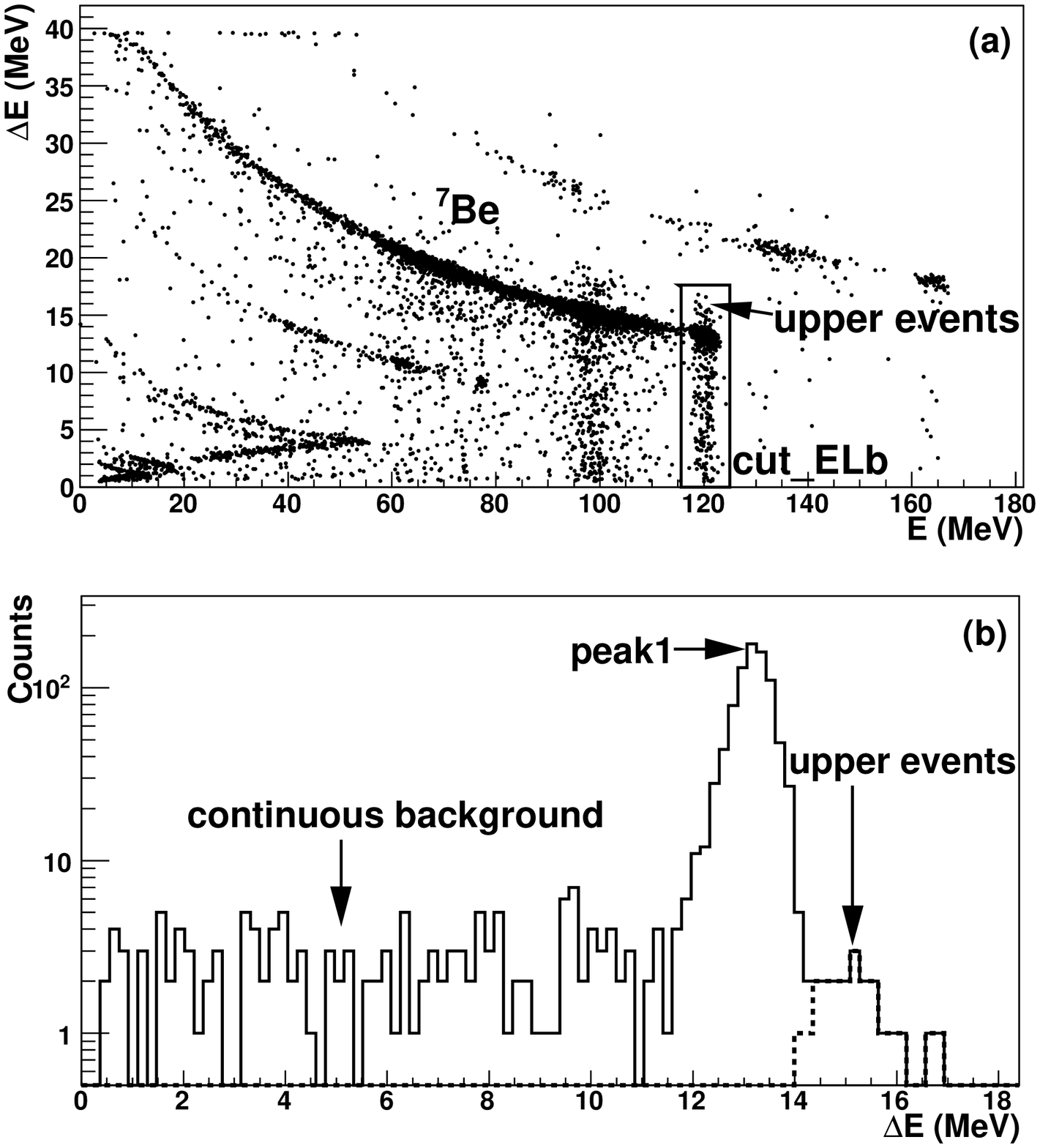}\\
  \figcaption{\label{fig_bdee} (a) $\Delta$E-E plot of the telescope. $\Delta$E is from one exit face strip of the DSSSD, E is from the SSD. The graphical cut named cut\_ELb encircled some events used for the analysis. Some events with higher $\Delta$E energies in the graphical cut cut\_ELb are marked as upper events. (b) Solid line is the $\Delta$E spectrum obtained by projecting the selected $^{7}$Be events in the graphical cut cut\_ELb to the Y-axis, and dashed line is the same $\Delta$E spectrum but some restrictive graphical cuts from junction side strips as shown in Fig.\ref{fig_chsh}(a) are imposed. See context for details.}
\end{center}

\subsection{Analysis of the online test results}
Fig.\ref{fig_chsh} shows the CS effect of the DSSSD working in transmission model. CS events are observed on both junction and ohmic side strips. These events are induced by those particles penetrating the corresponding interstrip gaps on both sides of the DSSSD\cite{int87}. So the position information of these events' can be recorded. And the energy can be got by summing the energy values of the two adjacent strips. The raised events in Fig.\ref{fig_chsh}(a) imply excess energy loss while particles punch through the junction side interstrip gaps. Except these raised events, the other CS events of junction side strips exhibit normal pattern that same as those constant line events in Fig.\ref{fig_chsh}(b). This means that only a part of those particles, which punching through the DSSSD from a position that very near to the junction side interstrip regions, will induce excess energy loss. The cause of these raised events will be interpreted infra.

The response of one entrance face strip of the DSSSD working in transmission mode is shown in Fig.\ref{fig_fdee}. The peak1 in Fig.\ref{fig_fdee}(b) is the elastic scattering peak of the beam on the target\cite{yang13}.

The continuous background in Fig.\ref{fig_fdee}(b) arise from the junction side CS events as shown in Fig.\ref{fig_chsh}(a). The ratio between the number of the counts in the continuous background and the counts of the whole $\Delta$E spectrum is $\sim$10\%. This is consistent with the geometrical ratio between the area of interstrip and strip regions of junction side, which also confirms that the CS events originate from the junction side interstip region.

The peak2 in Fig.\ref{fig_fdee}(b) represents the contaminant under each of the element belts in Fig.\ref{fig_fdee}(a). It is clear that the ICC effect of the junction side strip results in the appearance of the peak2, i.e. the contaminant events in Fig.\ref{fig_fdee}(a). The number of the counts in peak2, after subtract the included continuous background events, occupy $\sim$20\% of the counts of the whole $\Delta$E spectrum. This leads to a hypothesis that these events in the peak2 originate from the particles punching through the DSSSD from cross region 1 (see Fig.\ref{fig_dsssd}), which occupies $\sim$20\% of the area of the DSSSD. These ionising particles penetrating the DSSSD from cross region 1 will produce hole-electron pairs on its range. At the end of the track that near to the interstrip gap of the ohmic side, the holes have a considerable probability of recombination because of the weak electric field of this region and the long route to the cathode\cite{ef1,ef2} (junction side strip), however the electrons will be completely collected (by one or two adjacent strips of ohmic side) due to its high mobility. Correlation analysis is performed, i.e. the CS events of ohmic side strips (see Fig.\ref{fig_chsh}(b)) which originate from the interstrip region of ohmic side are used as restrictive cut to inspect the $\Delta$E spectrum of junction side strip. The result, as shown in Fig.\ref{fig_fdee}(b) with dashed lines, confirms that the events in peak2 of Fig.\ref{fig_fdee}(b) are homologous with the CS events of the exit face strips, i.e. these events originate from the particles punching through the DSSSD from cross region 1 (see Fig.\ref{fig_dsssd}). This validates the previous hypothesis on the origination of the peak2 in Fig.\ref{fig_fdee}(b), and also indicates that the ohmic side interstrip gaps results in a strong ICC effect on the hole collection of the junction side strips while the DSSSD works in transmission mode. For these ICC events of junction side strips, i.e. the events in peak2 of Fig.\ref{fig_fdee}(b), the accurate energy values from junction side strip are lost and can not be reconstructed. However the position information is recorded without ambiguity.

Fig.\ref{fig_bdee} shows the response of one exit face strip of the DSSSD working in transmission mode. The peak1 in Fig.\ref{fig_bdee}(b) is the elastic scattering peak\cite{yang13}. The continuous background in Fig.\ref{fig_bdee}(b) comes from the CS events as shown in Fig.\ref{fig_chsh}(b). The ratio of the number of the counts in the continuous background, i.e. the counts of CS events, to the counts of the whole spectrum is $\sim$15\%. This means that the CS effect of ohmic side is reduced as compared with the result of the test using $\alpha$ source (see Fig.\ref{fig_bsp}(b)), in which the ratio is consistent with the geometrical ratio between the area of ohmic side interstrip and strip regions, i.e. 20\%. The reduction of CS effect of ohmic side strips is related to the rise of the range of the particles in the DSSSD while illuminating its junction side\cite{must}.

The analysis of the online test result presents the energy and position information of the ICC and CS events, and ensures their validity as elastic scattering events\cite{yang13}. As a typical example, the position distribution of the elastic scattering events of the $^{7}$Be is shown in Fig.\ref{pdis}.

\begin{center}
  \includegraphics[width=9cm]{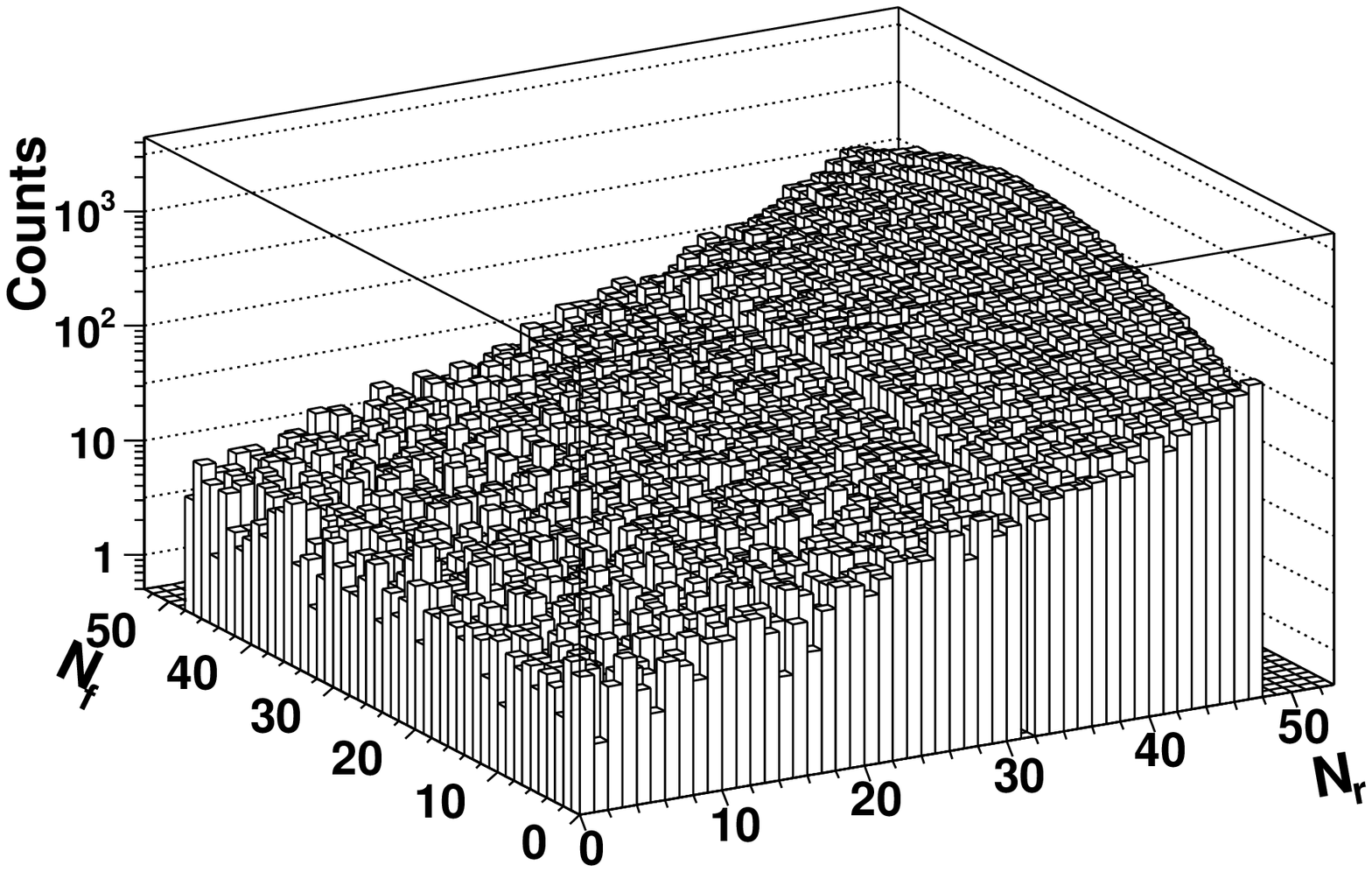}\\
  \figcaption{\label{pdis} 2D position distribution of the elastic scattering events. N$_{f}$ and N$_{r}$ is the strip number of the front and rear sides respectively.}
\end{center}

The dashed line in Fig.\ref{fig_bdee}(b) is the result of the correlation analysis of the upper events with the raised events in Fig.\ref{fig_chsh}(a). The result shows that the upper events are homologous with these raised events showing in Fig.\ref{fig_chsh}(a), which originates from the particles punching through the interstrip region of the junction side. This indicates that they have excess energy loss while the particles punch through the junction side interstrip gaps. The excess energy loss is related to the structure of the DSSSD in the interstrip region of junction side, i.e. the boundary structure of the junction side strips. It is found that, if the bias voltage of the DSSSD is removed, the raised events will become sunken as shown in Fig.\ref{fig_concave}. This implies that the depleted junction will extends underneath the interstrip gap of the junction side while appropriate bias voltage is imposed upon the DSSSD, and hence the holes deposited in this region will be collected by the two adjacent junction side strips.

\begin{center}
  \includegraphics[width=6cm]{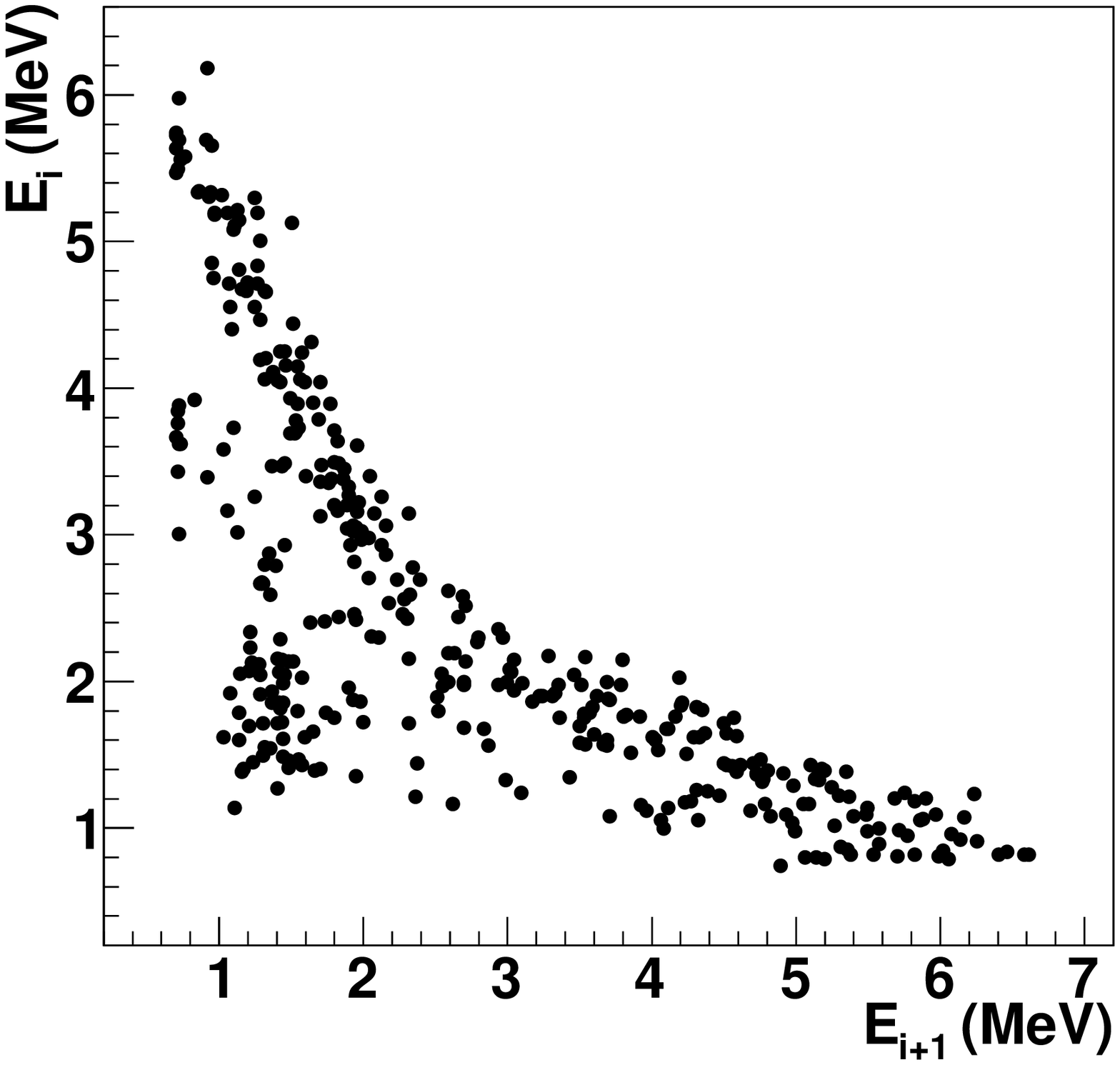}\\
  \figcaption{\label{fig_concave} Charge correlation between two adjacent junction side strips (strip i and i+1) of the DSSSD while the bias voltage is reduced.}
\end{center}

\section{Summery}
\label{con}
The performance of a DSSSD with effective area of $\sim$5$\times$5 cm$^2$ is reported. The effect of ICC and CS effects on the energy determination and position detection is evaluated by using short range $\alpha$-particles from an $^{241}$Am source and penetrating heavy ion beams illuminating the junction side of the DSSSD.
The analysis shows that the energy information of CS events can be obtained by summing the energy signal of the two adjacent strips. And the energy information of ICC events is deficient, and can not be reconstructed. However, the position information of these events can be recorded without ambiguity, benefiting from the knowledge of the origination of these events.
\\

\acknowledgments{We would like to express appreciation to Prof. Z.K. Li (Institute of Modern Physics, Chinese Academy of Science) who made helpful comments on the manuscript.}

\end{multicols}

\vspace{10mm}

\vspace{-1mm}
\centerline{\rule{80mm}{0.1pt}}
\vspace{2mm}

\begin{multicols}{2}

%
%
%
%
%

\bibliographystyle{unsrt}
\bibliography{DSSSD}

\end{multicols}

\clearpage

\end{document}